\documentclass[prl,twocolumn]{revtex4}
\usepackage{graphicx}

\begin{document}

\def\K{{\bf{K}}}
\def\Q{{\bf{Q}}}
\def\Gbar{\bar{G}}
\def\tk{\tilde{\bf{k}}}
\def\k{{\bf{k}}}

\title{Synergistic Polaron Formation in the Hubbard-Holstein Model at Small 
Doping}

\author{Alexandru Macridin$^{1}$, Brian Moritz$^{1,2,3}$, M.\ Jarrell$^{1}$, Thomas Maier$^{4}$}
\address{
$^{1}$ University of Cincinnati, Cincinnati, Ohio, 45221, USA\\
$^{2}$ University of Waterloo, Waterloo, Ontario, N2L 3G1, Canada \\
$^{3}$ University of North Dakota, Grand Forks, North Dakota, 58202, USA\\
$^{4}$ Oak Ridge National Laboratory, Oak Ridge, Tennessee, 37831, USA}

\date{\today}

\begin{abstract}

  We study the effect of dynamical Holstein phonons on the physics of
  the Hubbard model at small doping using the dynamical cluster
  approximation on a $2\times2$ cluster. Non-local antiferromagnetic correlations are found
  to significantly enhance the electron-phonon coupling, resulting in
  polaron formation for moderate coupling strengths.  At finite
  doping, the electron-phonon coupling is found to strongly enhance
  the non-local spin correlations, indicating a synergistic interplay
  between the electron-phonon coupling and antiferromagnetic
  correlations.  Although it enhances the pairing interaction, the
  electron-phonon coupling is found to decrease the superconducting
  transition temperature, due to the reduction in the quasiparticle
  fraction.

\end{abstract}

\pacs{}
\maketitle


{\em{Introduction.}}  Phonons significantly modify the properties of
conventional metals, giving rise to phenomena such as
superconductivity and charge-density wave formation.  These effects
are understood through conventional theories.  However, the effect of
phonons on strongly correlated metals is less understood. This problem
demands investigation given the experimental evidence of
significant electron-phonon (EP) interactions in
correlated materials such as the cuprates and manganites
\cite{ep_cuprates,ep_manganites}.

Theoretical interest is motivated by the suggestion that electronic 
correlations greatly enhance the effective EP coupling~\cite{j_zhong_92}.  
In this scenario, the interaction of the holes with their antiferromagnetic 
(AF) background results in band narrowing which enhances the effective EP 
coupling, driving the system to the polaron regime even for moderate EP
couplings~\cite{j_zhong_92,p_prolovsek_06,mishchenko}.  Quantum Monte Carlo 
(QMC) calculations~\cite{huang:qmc}, for the Hubbard-Holstein (HH) model 
predict that the EP coupling enhances d-wave pairing in the large coupling 
regime.  Other calculations for the HH or closely related t-J models, 
including finite size
\cite{b_alder_97,j_riera_06,p_prolovsek_06,t_sakai_97} and dynamical
mean field approximation (DMFA)
calculations~\cite{sangiovanni:DMFA,capone}, suggest that phonons
contribute strongly to spin and charge ordering, pairing, and phase
separation.

In this letter we employ the dynamical cluster approximation
(DCA)~\cite{hettler:dca,maier:rev} to study the two dimensional HH
model at small doping. We investigate the role of the EP coupling on
its one and two particle properties. The DOS, charge susceptibility and
unscreened local moment indicate band narrowing and polaron formation
at moderate EP couplings. Since this implies reduced mobility of the
holes, antiferromagnetism is strongly enhanced at finite doping, but
not at half filling. We find that the effective AF exchange as well as
the pseudogap observed in both the single-particle spectra and the
uniform spin susceptibility are nearly independent of the EP
coupling. D-wave superconductivity is suppressed by the phonons due to
the reduction of quasiparticle weight.

The HH Hamiltonian reads
\begin{eqnarray}
H&=&-t \sum_{\langle ij\rangle\sigma}\ \left(c^\dagger_{i\sigma} c_{j\sigma} +
c^\dagger_{j\sigma} c_{i\sigma}\right) + U \sum_i n_{i\uparrow} n_{i\downarrow}
\label{Eq:Ham} \nonumber 
\\
&+& \sum_i \frac{p_i^2}{2M} + \frac{1}{2} M \omega_0^2x_i^2 + g  n_i x_i.
\end{eqnarray}
\noindent with a nearest-neighbor hopping $t$, an on-site Coulomb repulsion 
$U$,  dispersionless optical phonons with frequency $\omega_0$ and an 
on-site EP coupling $g$.  The dimensionless EP coupling is defined as 
$\lambda=2 g^2/(2M\omega_0^2 W)$ and represents the ratio of the 
single-electron lattice deformation energy $E_p=g^2/(2M\omega_0^2)$ to half 
of the electronic bandwidth  $W/2=4t$. We present results for small phonon 
frequency $\omega_0=0.3t$, $U=8t$ and values of $\lambda$ in the weak and 
intermediate coupling regime, $\lambda\le 0.7$.

To study the Hamiltonian (\ref{Eq:Ham}) we employ the DCA, a cluster 
mean-field theory which maps the original lattice model onto a
periodic cluster of size $N_c=L_c^2$ embedded in a self-consistent
host. Correlations up to a range $L_c$ are treated explicitly, while
those at longer length scales are described at the mean-field level.
With increasing cluster size, the DCA systematically interpolates
between the single-site DMFA result and the exact result, while
remaining in the thermodynamic limit.
Cluster mean field calculations on the simple Hubbard model are found 
to reproduce many of the features of the cuprates, including a 
Mott gap and strong AF correlations, 
d-wave superconductivity and pseudogap behavior~\cite{maier:rev}.

We solve the cluster problem using a quantum Monte Carlo (QMC)
algorithm~\cite{jarrell:dca} modified to perform the sum over both the
discrete field used to decouple the Hubbard repulsion, as well as the
phonon field $x$.  The space of configurations of the latter field is
significantly larger than the former, so the HH code
requires significantly more CPU time than required for the Hubbard
model.  Thus, present calculations are restricted to clusters of size
$N_c=4$.  The Maximum Entropy method~\cite{jarrell:mem} is employed to
calculate the real frequency spectra.

\begin{figure}[t]
\begin{center}
\includegraphics*[width=3.3in]{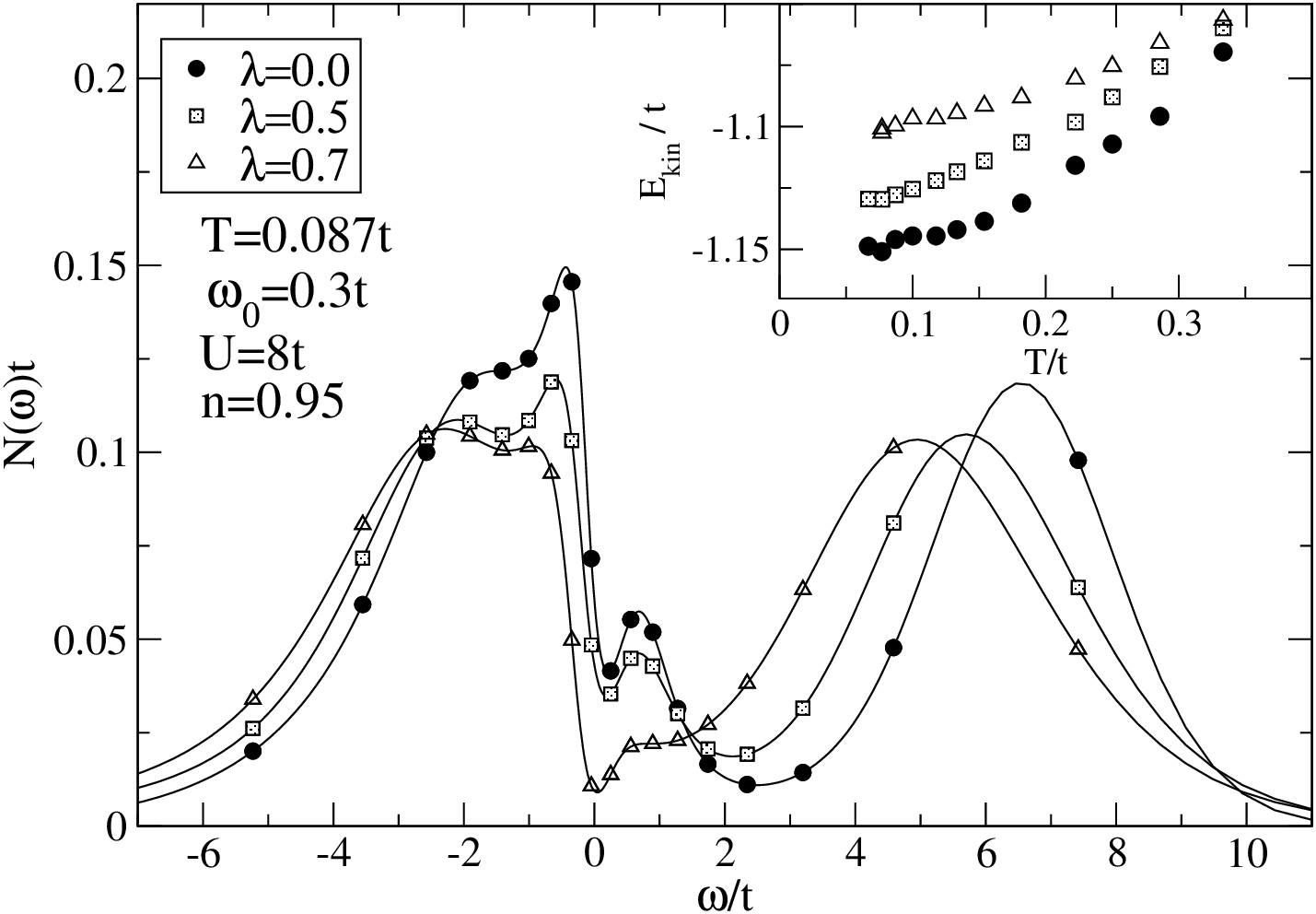}
\caption{Single-particle DOS for several values of $\lambda$.
 The pseudogap width is negligibly dependent
 on $\lambda$. Inset: the electronic kinetic energy, $E_{kin}=\sum_{k
    \sigma}\epsilon(k) n_{k,\sigma}$, versus T.  $N(0)$ and the kinetic 
    energy gain are suppressed with
  increasing $\lambda$.}
\label{fig:dos}
\end{center}
\end{figure}

{\em{Results.}} The electronic DOS at $5\%$ doping is plotted in
Fig.~\ref{fig:dos}.  With increasing $\lambda$ the DOS at the Fermi
energy, $N(0)$, is strongly suppressed. The pseudogap in the DOS, as
measured by the peak to peak distance, does not depend much on
$\lambda$. However, the peaks adjacent to the pseudogap are strongly
suppressed by the EP coupling. The EP coupling also reduces the
separation of the Mott peaks which is a measure of $U_{eff}$.  The
reduction depends on the phonon frequency (not shown) and for small
$\omega_0$ we find that it is much smaller than that expected in the
antiadiabatic limit ($U_{eff}=U-\lambda W$ when $\omega_0 \to \infty$).

The inset in Fig.~\ref{fig:dos} shows that the electronic kinetic
energy gain at low temperatures decreases with increasing
$\lambda$. The quasiparticle fraction, which will be discussed below
(see Fig.~\ref{fig:Tcsvslambda}), also falls with increasing
$\lambda$.  These results, together with the suppression of $N(0)$ and
the peaks adjacent to the pseudogap, which in previous calculations
were associated with quasiparticles~\cite{maier:rev}, are consistent
with a significant decrease in the effective electronic bandwidth and
a reduction of the coherent part of the electronic Green's function
due to EP coupling.

\begin{figure}[t]
\begin{center}
\includegraphics*[width=3.3in]{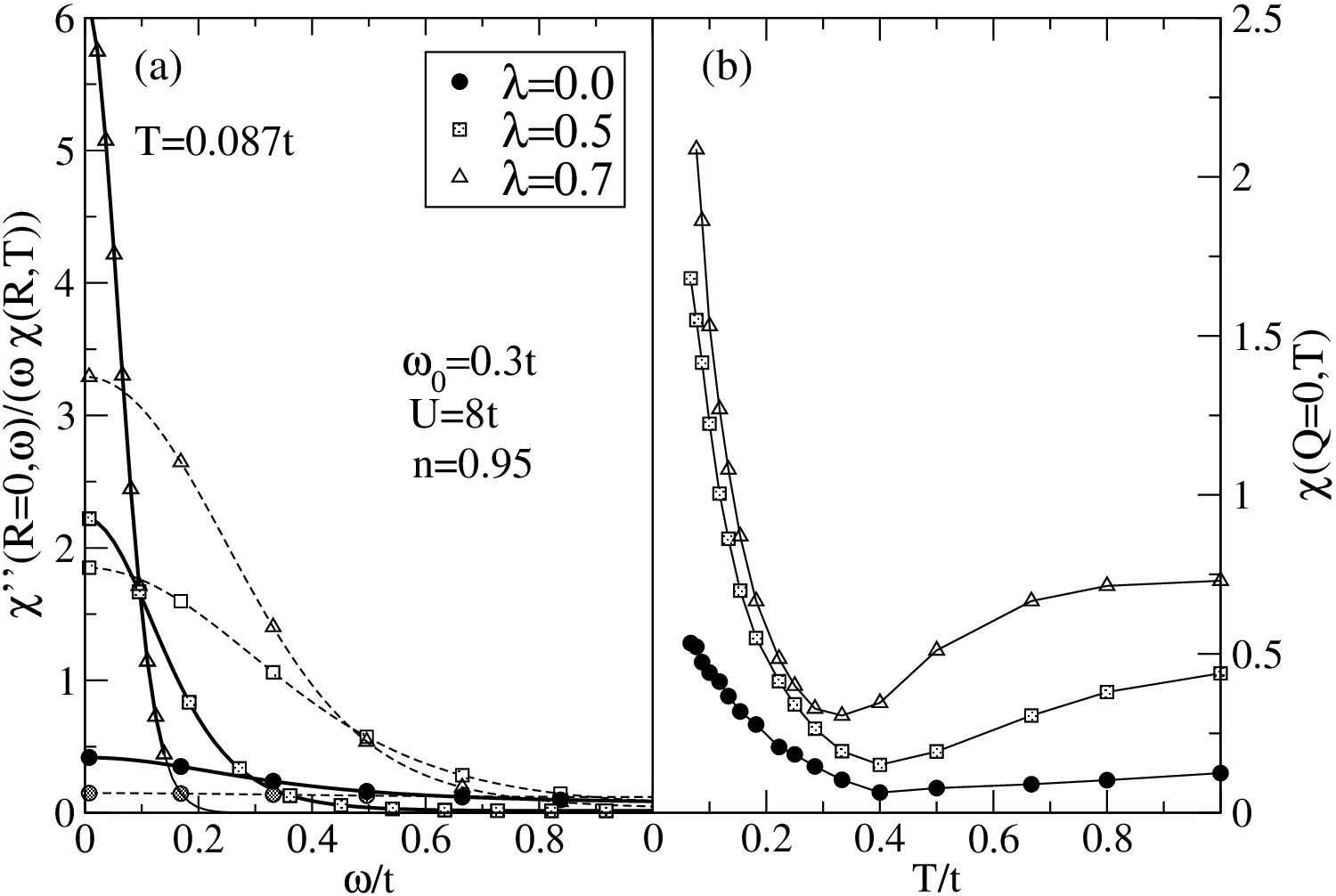}
\caption{(a) The local dynamic charge
  susceptibility for several values of $\lambda$.  The dashed (solid)
  lines are DMFA (DCA) results.  A narrow band of low energy charge
  excitations develops as $\lambda $ increases which indicates polaron
  formation.  (b) Static charge susceptibility (compressibility)
  versus $T$ for the same values of $\lambda$. The strong enhancement
  indicates the tendency to phase separation.}
\label{fig:chasus}
\end{center}
\end{figure}

The local dynamical charge susceptibility is plotted in 
Fig.~\ref{fig:chasus}(a). It develops a narrow peak with increasing
$\lambda$, suggesting the development of a nearly localized band of
polaron charge carriers. When dynamical phonons are considered 
there is no self-localization, making 
it difficult to
define a critical $\lambda_c$ for polaron formation.  For
$\lambda=0$ the width of the peak is of order of $t$, while for $\lambda \ge
0.5$ the width of the peak becomes smaller than the phonon frequency
$\omega_0$ which we take to indicate the crossover to a polaron
regime.  Note that the peak found in the DCA (solid lines) is much
narrower than that found by the DMFA (dashed lines) for the same
parameters~\cite{DMFAcaveat}.  This is consistent with the
argument~\cite{j_zhong_92} that the effective EP coupling is enhanced
by non-local AF correlations, which are present in the DCA but not in
the DMFA.  Polaron formation is also associated with phase separation,
since the electronic band is narrowed, i.e.\ $t_{eff}$ is reduced, and
the effective AF exchange $J_{eff}$, as we will show below,
essentially remains unchanged for small $\omega_0/t$.  An increased
$J_{eff}/t_{eff}$ leads to phase separation since the system can gain
more energy from spin exchange than it can from the kinetic energy of
the carriers~\cite{emerykivelson}. In accord with this argument, the
bulk charge susceptibility (or compressibility), shown in
Fig.~\ref{fig:chasus}(b), is strongly enhanced by the EP coupling 
(for larger values of $\lambda$ than those shown, it diverges at low
$T$) indicating a charge ordering transition (phase separation).

\begin{figure}[t]
\begin{center}
\includegraphics*[width=3.3in]{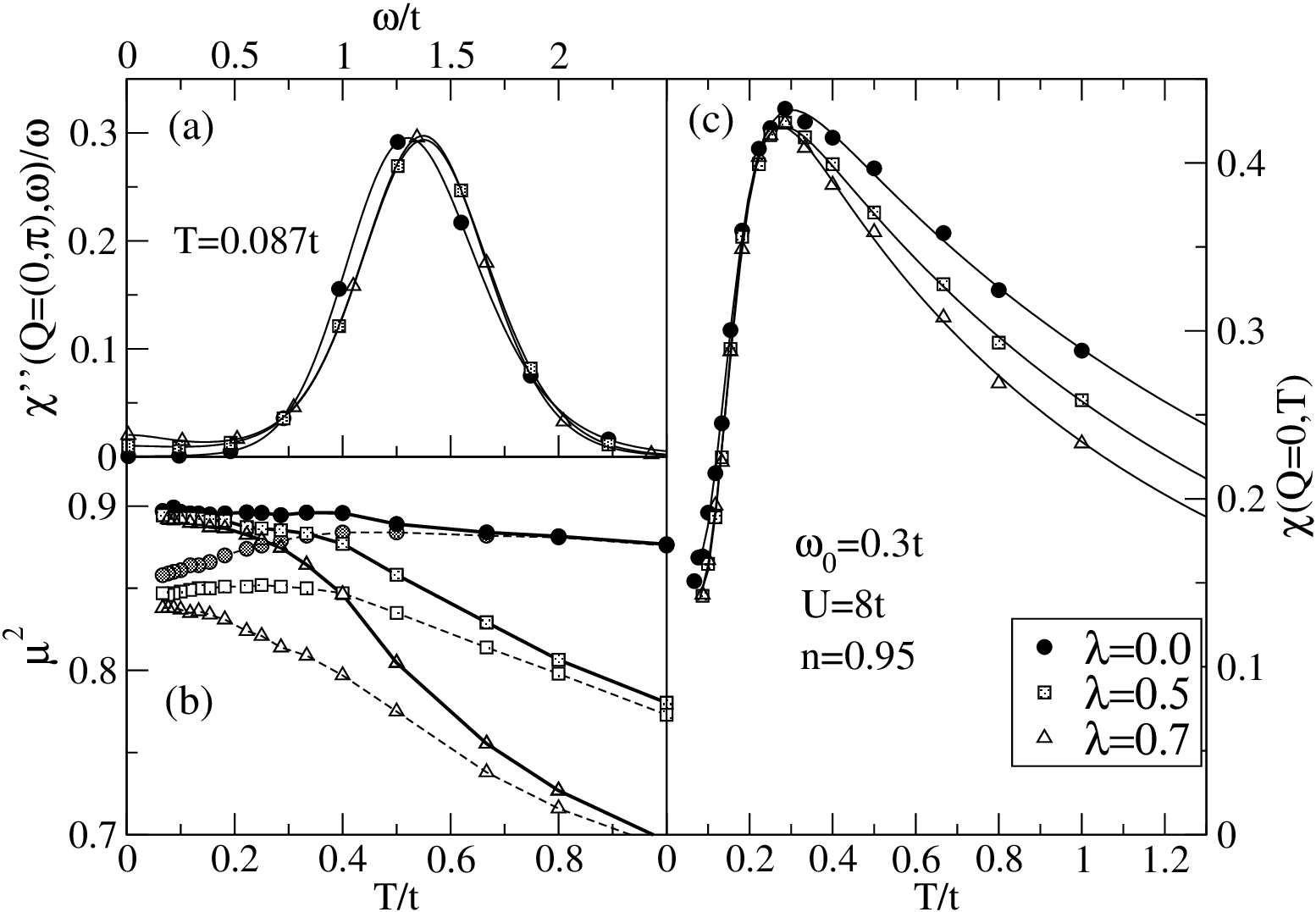}
\caption{(a) The dynamic cluster spin susceptibility with $Q=(\pi,0)$.
  The location of the peak is a measure of the effective spin exchange
  $J_{eff}$. (b) The unscreened moment
  $\mu^2=\langle(n_\uparrow-n_\downarrow)^2\rangle$ versus $T$,
  calculated with the DCA (solid lines) and the DMFA (dashed lines).
  The EP coupling strongly increases the temperature-dependence of
  $\mu^2$ but the low-temperature value is unaffected. (c) The bulk
  spin susceptibility versus T.}
\label{fig:spinsus}
\end{center}
\end{figure}

\begin{figure}[t]
\begin{center}
\includegraphics*[width=3.3in]{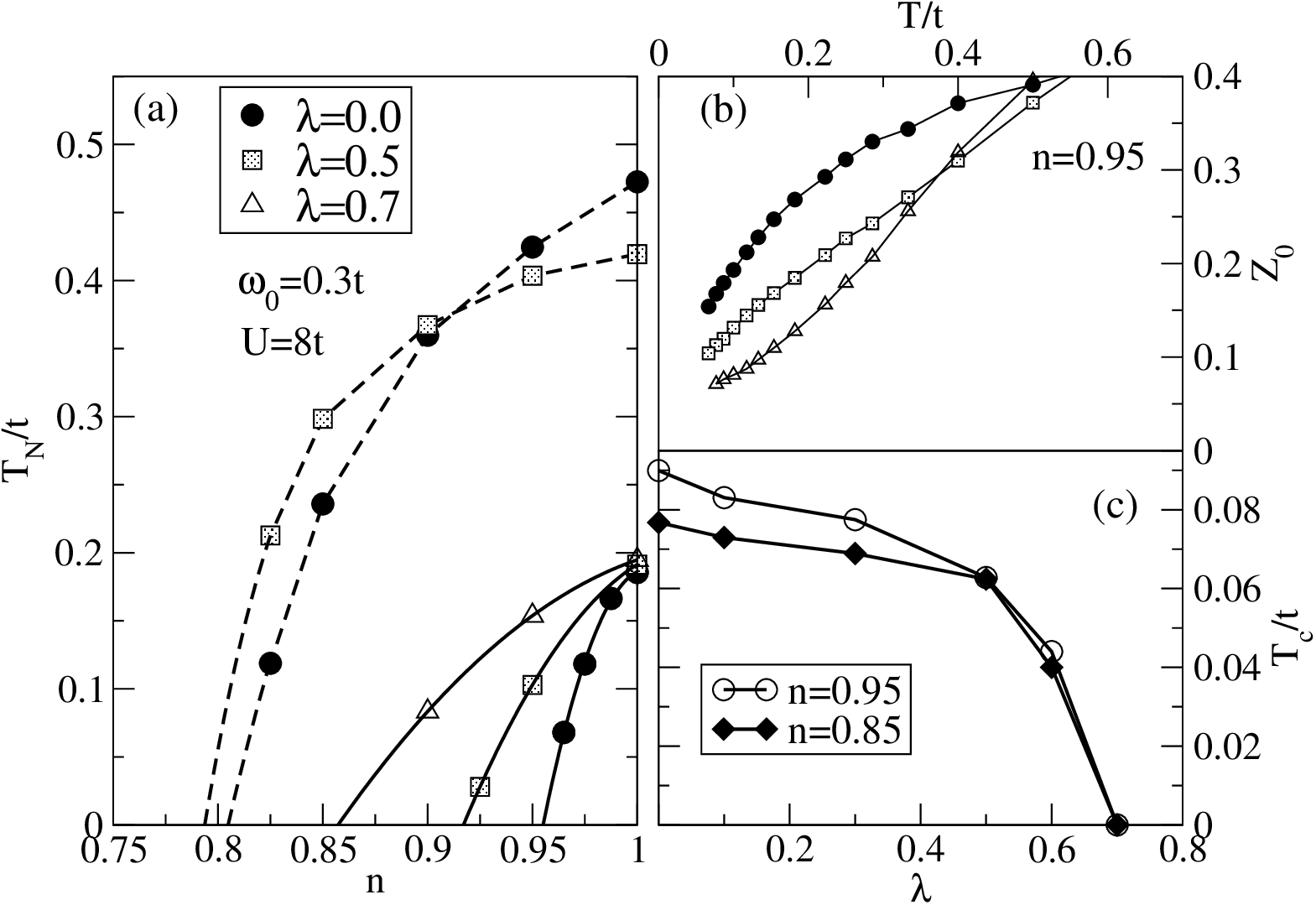}
\caption{ (a) $T_N$ versus
  filling for several values of $\lambda$ calculated with the DCA
  (solid lines) and the DMFA (dashed lines).  When $n=1$, the DCA
  $T_N$ increases slightly, consistent with a small increase in
  $J_{eff}$ shown in Fig.~\ref{fig:spinsus}(a).  At small doping,
  $T_N$ is enhanced by the EP coupling in contrast to the DMFA result
  where $T_N$ is suppressed.  (b) The Matsubara quasiparticle fraction
  $Z_0=1/(1- {\rm{Im}} \Sigma(K,\pi T)/\pi T)$ versus $T$ for several
  values of $\lambda$.  (c) Superconducting transition temperature
  versus $\lambda$ at $5\%$ and $15\%$ doping. }
\label{fig:Tcsvslambda}
\end{center}
\end{figure}

The dynamic cluster spin susceptibility is plotted in
Fig.~\ref{fig:spinsus}(a) for $Q=(0,\pi)$.  Due to remnant AF spin
correlations, it has a magnon peak at an energy
proportional to the effective exchange interaction
$J_{eff}$. The fact that
$J_{eff}$ increases very slowly with $\lambda$ is expected from
perturbation theory in $t$~\cite{macridin:bipolaron} and 
exact diagonalization results~\cite{ed:stephan} 
when $\omega_0$ is small.  $J_{eff}$ is
independent of $\lambda$ when $\omega_0\to 0$ since the lattice cannot 
respond during the virtual hopping between near neighbors
which lowers the energy of the singlet relative to the triplet.

The unscreened moment $\mu^2=\langle (n_\uparrow-n_\downarrow)^2\rangle$ 
is plotted in Fig.~\ref{fig:spinsus}(b). In the DCA calculation,
increasing $\lambda$ greatly increases the temperature dependence of
$\mu^2$, but at low temperatures,  $\mu^2$ is essentially
independent of $\lambda$. In the DMFA calculation, at the same
temperatures the EP coupling is found to significantly reduce $\mu^2$
due to the tendency of the phonons to reduce the on-site
correlations~\cite{sangiovanni:DMFA}. In the DCA, the reduction of the
effective hopping amplitude due to the larger effective EP coupling
(i.e. enhanced polaron formation) counteracts the reduction of on-site
correlations, thus leaving the low temperature unscreened moment
unaffected.

The bulk spin susceptibility is plotted in Fig.~\ref{fig:spinsus} (c).
Unlike the charge susceptibility, the low temperature bulk spin
susceptibility is essentially independent of $\lambda$.  When the bulk
susceptibility is divided by the unscreened moment $\mu^2$, the four
curves collapse upon each other (not shown), with small deviations
consistent with the slight increase of $J_{eff}$ with $\lambda$.  This
indicates that the temperature and $\lambda$ dependence of the bulk
susceptibility, is due predominantly to its effect on the unscreened
moment.  The emergence of the pseudogap is associated with the
suppression of the spin excitations and it is interesting to notice
the weak $\lambda$-dependence of both the low-T bulk spin
susceptibility and the pseudogap in the DOS. This suggests that the
pseudogap energy scale is $J_{eff}$. 

The AF transition temperature $T_N$ at half filling ($n=1$) is weakly
dependent on $\lambda$ as shown in Fig.~\ref{fig:Tcsvslambda}(a).  It
increases only slightly, consistent with the increase in $J_{eff}$
with $\lambda$ found in Fig.~\ref{fig:spinsus}(a).  However, when the
system is doped, $T_N$ increases dramatically with $\lambda$ since 
polaron formation suppresses the mobility of the holes that disrupts 
the AF order.  Hence, not only do the AF correlations enhance the EP 
coupling~\cite{j_zhong_92} and polaron formation~\cite{p_prolovsek_06}, 
but the EP coupling also strongly enhances antiferromagnetism in the 
doped model.  This establishes a synergistic cooperation between the 
EP coupling and AF correlations.  Note that at half filling and small 
dopings, unlike DCA, the DMFA predicts a reduction of $T_N$ associated 
with the decrease in $U_{eff}$.  However, like the DCA, at larger 
doping the AF order is enhanced in DMFA due to the decrease of charge 
carrier mobility, although the effect is not as pronounced.

It is unclear a priori if the EP coupling will enhance superconductivity.
For $U>6t$ the EP vertex at small
momentum transfers increases with increasing $U$, hence contributing
to a d-wave pairing interaction~\cite{huang:qmc}. However, the EP
coupling also 
suppresses $N(0)$ as well as the
Matsubara quasiparticle fraction $Z_0=1/(1- {\rm{Im}} \Sigma(K,\pi
T)/\pi T)$.   It is plotted in Fig.~\ref{fig:Tcsvslambda}(b) for different
$\lambda$ for $K=(0,\pi)$.  As $T\to 0$, $Z_0\to Z$ where $Z$ is the
quasiparticle fraction defined by the derivative of the real-frequency
self energy.  At low temperatures the quasiparticle fraction is
strongly reduced for large $\lambda$. The superconducting transition
temperature is plotted in Fig.~\ref{fig:Tcsvslambda}(c) versus
$\lambda$ at $5\%$ and $15\%$ doping.  The leading pairing instability
remains even-frequency d-wave.  $T_c$ falls with increasing $\lambda$,
indicating that the reduction of $N(0)$ and the quasiparticle fraction
outweighs the enhanced interaction. Alternatively, several authors have 
proposed that pairing in the cuprates could be driven by kinetic energy 
gain~\cite{maier:kinetic,sc_stripe}.  The suppression of the kinetic 
energy seen in Fig.~\ref{fig:dos} is consistent with the suppression 
of $T_c$ in such scenarios.  

While our results are valid at small $\omega_0$, qualitative
differences are expected when it increases.  However, note that in
real materials the phonon energy is even smaller than our value, e.g.\
in cuprates $\omega_0 \approx 0.10 t,~0.15 t$~\cite{ep_cuprates}.

Several experimental features in the cuprates
can be addressed with our results. The enhancement of the EP coupling
due to AF correlations can explain the large $\lambda \approx 1.5$ 
seen in the  photoemission spectra in the underdoped cuprates~\cite{ep_cuprates},
whereas density  functional calculations predict a moderate EP coupling
$\lambda \approx 0.4$~\cite{OKA}. The enhancement of the
charge susceptibility (Fig.~\ref{fig:chasus}) shows that EP
coupling is relevant for the charge inhomogeneities observed in
many high $T_c$ materials~\cite{inhomog}.

There is evidence in the literature that AF correlations can enhance
EP coupling, and mixed results regarding polaron formation and
pairing.  Exact diagonalization calculations with both
adiabatic~\cite{p_prolovsek_06} and
dynamic~\cite{t_sakai_97,j_riera_06} phonons, finite-size
QMC~\cite{huang:qmc} and diagrammatic QMC~\cite{mishchenko} all
suggest that AF correlations strongly enhance the effect of EP
coupling.  Those with dynamic~\cite{t_sakai_97} phonons find 
evidence that the phonons can enhance the equal time AF correlations.
However, these calculations have an extremely truncated
phonon Hilbert space that limits their ability to properly describe
polaron formation. Perhaps as a consequence, these calculations
suggest that Holstein phonons enhance d-wave superconductivity. 
Bipolaron calculations~\cite{macridin:bipolaron,manga} suggest 
enhanced formation of mobile intersite singlets for intermediate EP 
coupling.  However, the weak $\lambda$-dependence of the spin susceptibility,
Fig.~\ref{fig:spinsus} (c), excludes this scenario.
Bipolaron calculations, like exact diagonalization at small doping, 
are restricted to one or two carriers,  while the DCA addresses finite 
doping  in the thermodynamic limit.  In recent weak-coupling functional 
RG calculations, Holstein phonons were found to reduce the AF susceptibility, 
as well as pairing (but certain lower symmetry modes enhance 
pairing)~\cite{c_honnerkamp}.  However, these results neglect polaronic 
effects.

{\em{Conclusion.}}  We study the effect of Holstein phonons on the
properties of the two-dimensional Hubbard model using the DCA and QMC.  
At small doping we find that $N(0)$ and the quasiparticle fraction are 
strongly suppressed with increasing EP coupling.  This, together with 
a strongly increased charge susceptibility and the formation of a narrow 
band of charge excitations in the dynamic susceptibility, are taken as 
strong evidence for the formation of polarons.  The evidence is far 
stronger in the DCA than the DMFA results, consistent with the argument 
that non-local spin correlations strongly enhance the 
EP coupling~\cite{j_zhong_92}.  The spin properties captured by the DCA 
show that while the AF exchange $J_{eff}$ increases slightly, the 
primary effect of the EP coupling is to increases the temperature 
dependence of the unscreened moment $\mu^2$ while essentially
preserving its low-temperature value. Despite the reduction of $U_{eff}$ 
the phonons do not suppress the
low-$T$ moment, since the polarons suppress the hybridization of the
sites with their environment.   The AF transition temperature $T_N$ 
at half filling increases slightly, by an amount commensurate with the 
increase in $J_{eff}$.  However, at finite doping $T_N$ increases 
dramatically, due to the suppression of the itinerant holes mobility 
by polaron formation.  Thus not only do the AF correlations increase the 
effective EP coupling but also the EP interaction increases the AF 
correlations.  We find that both the magnitude of the pseudogap in the 
single particle DOS and the pseudogap signature in the bulk spin
susceptibility are not affected significantly by EP coupling,
suggesting that the pseudogap energy scale is $J_{eff}$.
Finally, although the EP coupling is expected to increase the d-wave
pairing interaction~\cite{huang:qmc}, we find that the superconducting
$T_{c}$ is actually suppressed due to the reduction of the quasiparticle
fraction $Z$ and $N(0)$.  However it is possible that couplings to 
other phonon modes with lower symmetry, may contribute an interaction 
which will enhance the d-wave $T_c$.  This will be a subject of future 
studies.

\acknowledgments We thank G.\ Sawatzky,  T.\ Devereaux, C.\ Honerkamp, 
R.\  Scalettar and H.-B.\ Sch\"uttler for useful discussions. This research 
was supported by NSF DMR-0312680, CMSN DOE DE-FG02-04ER46129, 
ONR N00014-05-1-0127 and NSF SCI-9619020 through resources provided by the 
San Diego Supercomputer Center.  TM acknowledges the Center for Nanophase 
Materials Sciences, sponsored by the Division of Scientific User Facilities, 
U.S. Department of Energy. BM acknowledges the UND Computational Research 
Center, supported by NSF EPS-0132289 and EPS-0447679, and the hospitality 
of the Pacific Institute of Theoretical Physics.


\begin{thebibliography}{99}
\bibitem{DMFAcaveat} At low $T$ the DMFA DOS and dynamic charge susceptibility develop phonon 
features at $\omega\approx \omega_0$ not found with the DCA at the temperatures studied.
\end{thebibliography}
\end{document}